\documentclass[onecolumn,showpacs,preprintnumbers]{revtex4}
\usepackage{dcolumn}
\usepackage{bm}
\usepackage{amsmath}
\usepackage{graphicx}
\usepackage{mathrsfs}
\usepackage{amssymb}
\usepackage{amsfonts}
\usepackage{indentfirst}
\begin{document}
\draft
\title{Entanglement of three cavity fields via resonant interactions with dressed
three-level atoms}
\author{Jinhua Zou{%
Author for correspondence. jhzou@yangtzeu.edu.cn}}
\address{College of Physical Science and Technology, Yangtze University, Jingzhou, 434023, China }

\begin{abstract}
In this paper we show that three cavity fields can be entangled when they
are tuned on resonance with an ensemble of dressed three-level atoms. The
master equation for the three cavity modes is derived by using atomic
dressed states and the inseparability of the three output cavity modes is
described by using a sufficient criterion proposed by van Loock and
Furusawa. The physical cause is the atomic coherence effects, by which the
quantum correlations are created in the field dynamics.

\end{abstract}

\maketitle

\section{\protect\smallskip Introduction}

Atomic coherence lies in the center of many novel effects in quantum optics
and laser physics. Electromagnetically induced transparency [1,2], coherent
population trapping [2], Hanel-effect laser [3] and quantum beat laser [4]
are such examples. Besides these, the correlation between the photons can
also be induced by atomic coherence [5-12]. One such example is the
generation of squeezed light in a three-level cascade laser using atomic
coherence [5-8]. The atomic coherence can be created by preparing the atoms
initially in a coherent superposition state of the two states which are
dipole-forbidden [5-8] or driving the two states by a strong coherent field
[9-11] or Raman coupling the two states through the third auxiliary atomic
states [12]. For two-photon correlated-spontaneous-emission laser with
injected atomic coherence, it exhibits complete spontaneous-emission noise
quenching and phase squeezing simultaneously [5]. It has also been pointed
out that atomic coherence in a two-photon correlated emission laser system
can be used to generate a macroscopic two-mode entangled state and this
system can be treated as an entanglement amplifier [12].

Recently, the topic of continuous-variable entanglement has attracted much
attention as it is the base of all branches of quantum information and
communication protocols [13]. Among various entanglement generation schemes,
entanglement induced by atomic coherence has been extensively researched
[14-16]. For a nondegenerate three-level cascade laser with a subthreshold
nondegenerate parametric oscillator coupled to a vacuum reservoir, the
entanglement and squeezing for the two cavity modes in this combined system
is induced by the injected atomic coherence [14]. In a two-mode single-atom
laser with the atomic coherence exhibited by two classical laser fields,
entanglement between two field modes is demonstrated [15]. Later, it was
shown that in a three-level $\Lambda $ or V atomic system with two classical
driving fields and two cavity modes coupling corresponding transitions, by
exploring the two-channel interaction mechanism and using the
squeeze-transformed modes, continuous-variable entanglement between the two
modes is obtained and the best achievable entangled state approaches the
original EPR state [16]. The above work has mainly been confined to
two-partite systems.

With the progress in continuous-variable entanglement, the generation of
more than two partite entanglement has been paid much attention as it may be
the key ingredient for advanced multiparty quantum communication such as
quantum teleportation network [17], telecloning [18] and controlled dense
coding [19]. Among various generation schemes for tripartite systems, few
work has been done to generate tripartite entanglement using atomic
coherence. Most recently, a scheme to generate three-mode-entangled light
fields via the interaction between the four-level atoms and the cavity has
been proposed [20]. Three cavity modes are generated through three
successive transitions in the four-level cascade atoms. In addition to the
cavity modes, two strong classical fields drive a pair of two-photon
transitions in the four-level atoms. They show that the entanglement could
only be obtained in a short time as all the mean photons are amplified as
time elapses. Thus at steady time, the entanglement does not exist.

In this paper, we present a scheme to generate tripartite entanglement for
three cavity modes via the interaction for the three-level lambda atoms with
the three cavity modes and two classical fields. As the classical fields are
strong, the effective interaction is resonant interaction in the
dressed-state picture. We deduce the master equation of the three cavity
modes by means of the atomic dressed states and linear theory. The
sufficient inseparability criterion for continuous-variable entanglement is
used to demonstrate the entanglement properties of the three cavity modes
and the results show that our system can be used as a source to generate
tripartite entangled light even at steady state.

It is should be noted that, up to now many schemes have been proposed to
generate tripartite entanglement using linear optics or nonlinearities [17,
21-26]. It was theoretically predicted that using single-mode squeezed state
and linear optics, a truly N-partite entangled state can be generated [17].
Later, a continuous-variable tripartite entangled state was experimental
realized by combing three independent squeezed vacuum states [21]. At first,
the production of continuous-variable tripartite entanglement was presented
by mixing squeezed beams on unbalanced beamsplitters [21,22]. Recently,
generation of tripartite entanglement are focused on using cascade nonlinear
interaction in an optical cavity [23-25] or in a quasiperiodic superlattice
[26]. Among the latter are systems using parametric down-conversion with sum
frequency generation [23,25,26] or using single nonlinearity [24]. During
these nonlinear processes, the cavity modes couple with each other directly.
As these nonlinear processes are related to the higher-order polarization,
the efficiency of these processes are relatively small compared with the
processes related to linear polarization. In this way, these nonlinear
processes are not the best choice for the generation of high efficiency
tripartite entangled states.

Compared with the schemes based on the nonlinear processes [23-26], our
scheme is more effective as the generation process is resonant interaction
in the dressed states and it is only related to linear polarization. What's
more, the linear process provides much more parameters to choose than that
of the nonlinear processes, as the atomic parameters can be varied. Compared
with the scheme in Ref. [20], our scheme can provide steady state tripartite
entanglement while the entanglement produced in scheme [20] is just kept in
a quite limited time. And in Ref. [20], they use four-level cascade atomic
system and the effective processes in the dressed states of the driving
fields are all two-photon transitions. High excited states are involved in
their scheme. In our scheme we use three-level $\Lambda $ atomic system, and
the effective processes in the dressed states of the driving fields are all
single-photon transitions. When we take into the account of the atomic
spontaneous emission, their schemes seems to have more obstacles than ours.

The paper is organized as follows. In Sec. II, we discuss the essential
ingredients of the model and deduce the density-matrix equation for the
cavity fields in a dressed-state picture. In Sec. III, we present the output
correlation spectra by solving the equations of the cavity fields and
analyze the output tripartite continuous-variable entanglement
characteristics by using a sufficient criterion proposed by van Loock and
Furusawa. In Sec. IV, we give a brief conclusion.

\section{Model and equation}

We consider $N$ three-level lambda-type atoms in a three-mode cavity as
shown in Fig. 1(a). Two laser fields of frequencies $\omega _{l1,l2}$ drive
the transitions $|1,2\rangle \leftrightarrow $ $|3\rangle $, respectively.
Two cavity modes $a_{1,2}$ of frequencies $\omega _{c1,c2}$ couple the
atomic transition $|1\rangle \leftrightarrow $ $|3\rangle $, while the
cavity mode $a_{3}$ with frequency $\omega _{c3}$ couples the transition $%
|2\rangle \leftrightarrow $ $|3\rangle $. $\gamma _{l}$ ($l=1,2$) are the
atomic decay rates from level $|3\rangle $ to levels $|1,2\rangle $ and $%
\kappa _{l}$ ($l=1,2,3$) are the cavity loss rates. For simplicity, we
assume that $\gamma _{1}=\gamma _{2}=\gamma $ and $\kappa _{1}=\kappa
_{2}=\kappa _{3}=\kappa $. The three cavity modes are assumed to be in their
vacuum state initially. In the frame of the frequencies of the laser fields
and under the dipole and the rotating-wave approximations, the total
Hamiltonian is
\begin{figure}[tbph]
\centering
\includegraphics[width=13 cm]{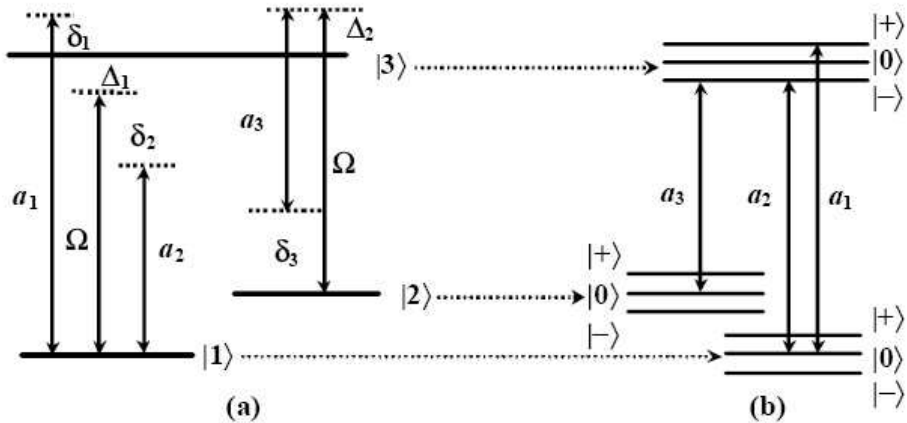}
\caption{(a) Atomic energy level scheme and the coupling of the cavity
fields and the classic fields. (b) Equivalent resonant transitions in the
picture dressed by the classical fields.}
\label{fig1}
\end{figure}
\begin{eqnarray}
H &=&H_{1}+H_{2}+H_{3},  \nonumber \\
H_{1} &=&\sum_{j=1}^{3}\hbar \delta _{j}a_{j}^{\dagger }a_{j},  \nonumber \\
H_{2} &=&-\hbar \Delta \left( \sigma _{11}-\sigma _{22}\right) +\hbar \Omega
\left( \sigma _{31}+\sigma _{32}+H.c.\right) ,  \label{1} \\
H_{3} &=&i\hbar g\left( a_{1}\sigma _{31}+a_{2}\sigma _{31}+a_{3}\sigma
_{32}\right) +H.c.,  \nonumber
\end{eqnarray}
H.c. symbols the Hermitian conjugate. $H_{1}$ denotes the free energy for
three cavity fields, $H_{2}$ describes the interaction of the laser fields
with the atoms, and $H_{3}$ indicates the interaction of the cavity fields
with the atoms. $\sigma _{jk}=|j\rangle \langle k|$ ($j,k=1,2,3$) are atomic
dipole operators for $j\neq k$ and atomic projection operators for $j=k$.
The cavity detunings are defined as $\delta _{j}=\omega _{cj}-\omega _{l1}$ (%
$j=1,2$), and $\delta _{3}=\omega _{c3}-\omega _{l2}$. The detunings of the
laser fields are defined as $\Delta _{j}=\omega _{3j}-\omega _{lj}$ ($j=1,2$%
), where $\omega _{31}$ and $\omega _{32}$ are the resonance frequencies of
transitions $|1,2\rangle \leftrightarrow $ $|3\rangle $. We have assumed
equal coupling coefficients $g$ for three cavity modes, equal Rabi frequency
$\Omega $ for the two laser fields, and opposite detunings of the two laser
fields $\Delta _{1}=-\Delta _{2}=\Delta $.

We assume that the laser fields are much stronger than the cavity fields,
i.e., $\Omega \gg g\langle a_{l}\rangle $, ($l=1,2,3)$. The laser fields can
be viewed as dressing fields for the atoms. Therefore, by diagonalizing the
Hamiltonian $H_{2}$, we find the so-called semiclassical dressed states as
\begin{eqnarray}
|0\rangle &=&-\frac{c}{\sqrt{2}}|1\rangle +\frac{c}{\sqrt{2}}|2\rangle
+s|3\rangle ,  \nonumber \\
|+\rangle &=&\frac{1+s}{2}|1\rangle +\frac{1-s}{2}|2\rangle +\frac{c}{\sqrt{2%
}}|3\rangle ,  \label{2} \\
|-\rangle &=&\frac{1-s}{2}|1\rangle +\frac{1+s}{2}|2\rangle -\frac{c}{\sqrt{2%
}}|3\rangle ,  \nonumber
\end{eqnarray}
where $c=\frac{\sqrt{2}\Omega }{d}$, $s=-\frac{\Delta }{d}$, and $d=\sqrt{%
\Delta ^{2}+2\Omega ^{2}}$.

Now, we use the Hamiltonian $H_{0}=\hbar d\left( \sigma _{++}-\sigma
_{--}\right) +H_{1}$ to perform the unitary dressing transformation. By
choosing the cavity detunings as $\delta _{1}=d=-\delta _{2}=-\delta _{3}$,
and neglecting the fast-oscillating terms such as $e^{\pm i2dt}$, we obtain
the resonant interaction Hamiltonian as
\begin{equation}
V=ig\hbar \left( c_{3}a_{1}^{\dagger }+c_{2}a_{2}+c_{1}a_{3}\right) \sigma
_{0+}+ig\hbar \left( c_{1}a_{1}-c_{3}a_{2}^{\dagger }+c_{3}a_{3}^{\dagger
}\right) \sigma _{0-}+H.c.,  \label{3}
\end{equation}
where $c_{1}=\frac{1}{2}s(1-s)$, $c_{2}=\frac{1}{2}s(1+s)$, and $c_{3}=\frac{%
1}{2}c^{2}$. The resonant transitions in the dressed states are shown in
Fig. 1(b).

The master equation for the cavity modes is obtained by using the usual
approach [2], starting from $\frac{d}{dt}\rho =-\frac{i}{\hbar }\left[
V,\rho \right] {\cal +L}_{a}\rho +{\cal L}_{c}\rho $, where ${\cal L}%
_{c}\rho =\frac{_{\kappa }}{2}\sum_{l=1}^{3}\left( 2a_{l}\rho a_{l}^{\dagger
}-a_{l}^{\dagger }a_{l}\rho -\rho a_{l}^{\dagger }a_{l}\right) $ and ${\cal L%
}_{a}\rho $ describes the atomic decay in the dressed states picture and its
expression is very complicated. The detailed form of atomic decay term $%
{\cal L}_{a}\rho $ is given in Appendix A. The master equation for the
cavity modes is obtained by tracing out the atomic states, which gives $%
\frac{d}{dt}\rho _{c}=g\left( c_{3}a_{1}^{\dagger
}+c_{2}a_{2}+c_{1}a_{3}\right) \rho _{+0}+g\left(
c_{1}a_{1}-c_{3}a_{2}^{\dagger }+c_{3}a_{3}^{\dagger }\right) \rho _{-0}+H.c$%
, where $\rho _{jk}=tr_{atom}(\sigma _{kj}\rho )$ ($j,k=0,+,-$). As the
atomic variables vary much faster than the cavity fields, it is possible to
express $\rho _{jk}=tr_{atom}(\sigma _{kj}\rho )$ ($jk=+0,-0,0+,0-$) in
terms of $\rho _{c}$, $a_{l}$ and $a_{l}^{\dagger }$ ($l=$1-3) from the
quasi-steady-state solution of the coupled equations for $\rho
_{jk}=tr_{atom}(\sigma _{kj}\rho )$ ($jk=+0,-0,0+,0-$). By using $\rho
_{jj}\simeq \rho _{jj}^{s}\rho _{c}$ ($j=0,+,-$) and $\rho _{+-}^{s}\simeq 0$%
, where ``s'' implies the steady-state solutions of the density matrix
equations in the dressed state picture without the quantum fields $a_{l}$
and $a_{l}^{\dagger }$ ($l=$1-3). The steady state populations is obtained
as $\rho _{00}^{s}=\frac{c^{4}}{1+3s^{4}}$ and $\rho _{++}^{s}=\rho
_{--}^{s}=\frac{s^{2}\left( 1+s^{2}\right) }{1+3s^{4}}$. The master equation
for the cavity modes is obtained as
\begin{eqnarray}
\frac{d}{dt}\rho _{c} &=&\sum_{l=1}^{3}\left\{ A_{ll}\left[ a_{l}^{\dagger
},\rho _{c}a_{l}\right] -\left( B_{ll}+\frac{\kappa _{l}}{2}\right) \left[
a_{l}^{\dagger },a_{l}\rho _{c}\right] \right\}  \nonumber \\
&&+\sum_{l=2}^{3}\left\{ A_{1l}\left[ a_{1}^{\dagger },\rho
_{c}a_{l}^{\dagger }\right] -B_{1l}\left[ a_{1}^{\dagger },a_{l}^{\dagger
}\rho _{c}\right] +A_{l1}\left[ a_{l}^{\dagger },\rho _{c}a_{l}^{\dagger
}\right] -B_{l1}\left[ a_{l}^{\dagger },a_{1}^{\dagger }\rho _{c}\right]
\right\}  \label{4} \\
&&+\sum_{l,k=2;l\neq k}^{3}\left\{ A_{lk}\left[ a_{l}^{\dagger },\rho
_{c}a_{k}\right] -B_{lk}\left[ a_{l}^{\dagger },a_{k}\rho _{c}\right]
\right\} +H.c..  \nonumber
\end{eqnarray}
The explicit expressions for $A_{lk}$ and $B_{lk}$ ($l,k=1,2,3$) are given
in Appendix B. Here the terms $A_{ll}$ ($l$=1-3) and $B_{ll}$ ($l$=1-3)
represent the gain term and the absorption of mode $a_{l}$, respectively.
And the terms $A_{lk}$ and $B_{lk}$ ($l\neq k$) represent the coupling
between the two modes $a_{l}$ and $a_{k}$, and we will show that these
quantities are responsible for entanglement among three cavity fields. It is
easy to see that without these coupling terms between different cavity
fields, the quantum correlation can not be introduced among the three cavity
modes. Thus entanglement among the three cavity fields is attributed to the
atomic coherence created through the interaction between the fields and the
atoms.

\section{Correlation spectra}

The master equation (4) enables us to derive equations of motion for the
cavity modes:
\begin{eqnarray}
\tau \frac{d}{dt}a_{1}^{\dagger } &=&\left( A_{11}-B_{11}-\frac{\kappa _{1}}{%
2}\right) a_{1}^{\dagger }+\left( A_{12}-B_{12}\right) a_{2}+\left(
A_{13}-B_{13}\right) a_{3}+\sqrt{\kappa _{1}}a_{1}^{\dagger in},  \nonumber
\\
\tau \frac{d}{dt}a_{2} &=&\left( A_{21}-B_{21}\right) a_{1}^{\dagger
}+\left( A_{22}-B_{22}-\frac{\kappa _{2}}{2}\right) a_{2}+\left(
A_{23}-B_{23}\right) a_{3}+\sqrt{\kappa _{2}}a_{2}^{in},  \label{5} \\
\tau \frac{d}{dt}a_{3} &=&\left( A_{31}-B_{31}\right) a_{1}^{\dagger
}+\left( A_{32}-B_{32}\right) a_{2}+\left( A_{33}-B_{33}-\frac{\kappa _{3}}{2%
}\right) a_{3}+\sqrt{\kappa _{3}}a_{3}^{in},  \nonumber
\end{eqnarray}
where $\tau $ is the round-trip time of light in the cavity and assumed to
be the same for three cavity modes. $a_{j}^{in}$ and $a_{j}^{\dagger in}$ ($%
j=$1-3) are annihilation and creation operators of the input fields to the
cavity. This is a set of linear equations. In order to solve this equation,
we use the Fourier transformation and the boundary conditions at the mirror
between the output quantities and the input quantities $%
a_{j}^{in}+a_{j}^{out}=\sqrt{\kappa _{j}}a_{j}$ ($j=$1-3) to obtain the
equation in the frequency domain as
\begin{equation}
a^{out}\left( \omega \right) =-\left( I+BD_{0}^{-1}B\right) a^{in}\left(
\omega \right) ,  \label{6}
\end{equation}
where
\begin{equation}
\begin{tabular}{l}
$a^{out}\left( \omega \right) =\left( a_{1}^{\dagger out}\left( -\omega
\right) ,a_{2}^{out}\left( \omega \right) ,a_{3}^{out}\left( \omega \right)
\right) ^{T},$ \\
$a^{in}\left( \omega \right) =\left( a_{1}^{\dagger in}\left( -\omega
\right) ,a_{2}^{in}\left( \omega \right) ,a_{3}^{in}\left( \omega \right)
\right) ^{T},$%
\end{tabular}
\label{7}
\end{equation}
\begin{equation}
D_{0}=\left(
\begin{array}{ccc}
A_{11}-B_{11}-\frac{\kappa _{1}}{2}-i\omega \tau & A_{12}-B_{12} &
A_{13}-B_{13} \\
A_{21}-B_{21} & A_{22}-B_{22}-\frac{\kappa _{2}}{2}-i\omega \tau &
A_{23}-B_{23} \\
A_{31}-B_{31} & A_{32}-B_{32} & A_{33}-B_{33}-\frac{\kappa _{3}}{2}-i\omega
\tau
\end{array}
\right) ,  \label{8}
\end{equation}
\begin{equation}
B=\left(
\begin{array}{lll}
\sqrt{\kappa _{1}} & 0 & 0 \\
0 & \sqrt{\kappa _{2}} & 0 \\
0 & 0 & \sqrt{\kappa _{3}}
\end{array}
\right) ,\text{\qquad }I=\left(
\begin{array}{lll}
1 & 0 & 0 \\
0 & 1 & 0 \\
0 & 0 & 1
\end{array}
\right) .  \label{9}
\end{equation}
where T symbols the matrix transpose.
\begin{figure}[tbph]
\centering
\includegraphics[width=15 cm]{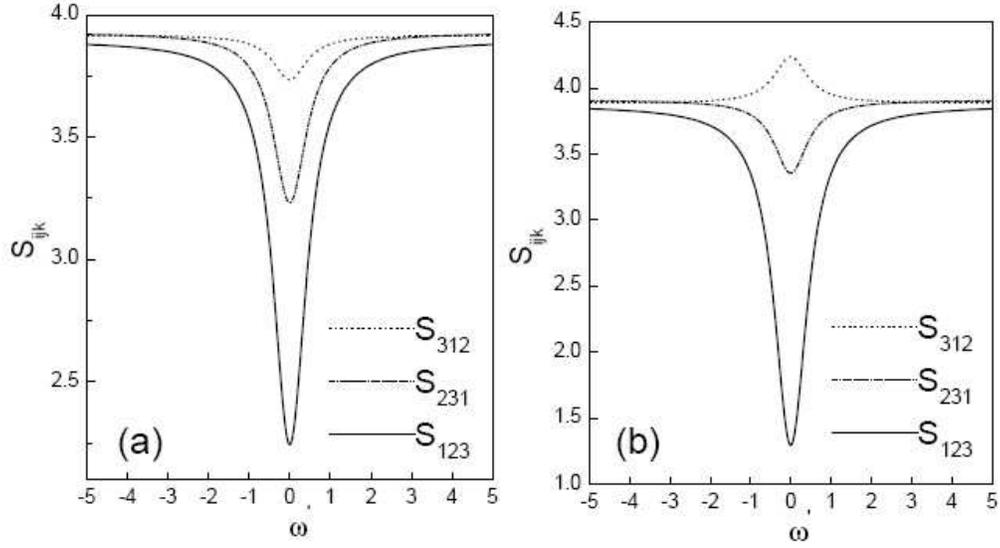}
\caption{The quantum correlations spectra $S_{123}^{out}\left( \omega
^{\prime }\right) $, $S_{231}^{out}\left( \omega ^{\prime }\right) $ and $%
S_{312}^{out}\left( \omega ^{\prime }\right) $ versus the normalized
analyzing frequency $\omega ^{\prime }$ are plotted for (a) $\Delta =5$ and
(b) $\Delta =10$ by solid, dashed and dotted line, respectively. The other
parameters are $\Omega =35$, $g^{2}N=10$, $\gamma =1$ and $\kappa =0.1$.}
\label{fig2}
\end{figure}
\begin{figure}[tbph]
\centering
\includegraphics[width=15 cm]{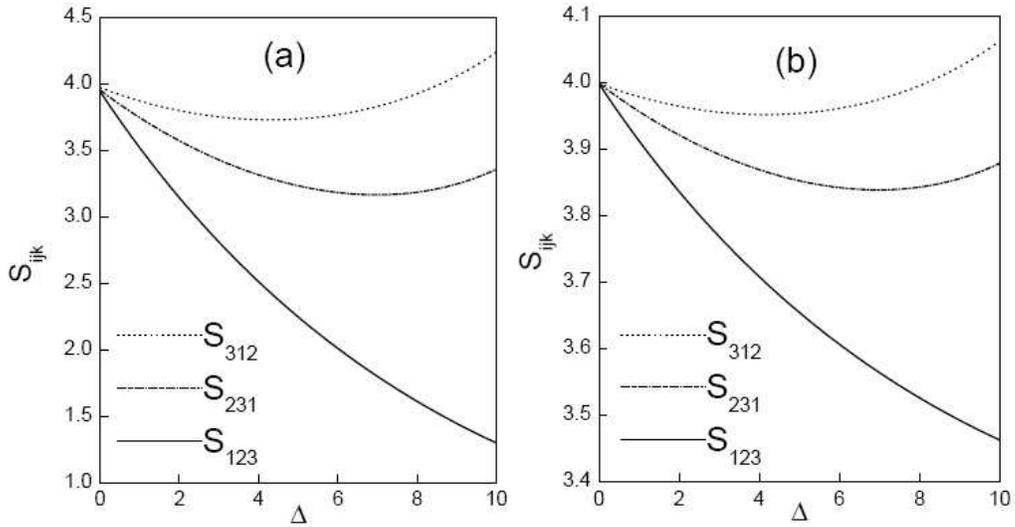}
\caption{The quantum correlations spectra $S_{123}^{out}\left( \omega
^{\prime }\right) $, $S_{231}^{out}\left( \omega ^{\prime }\right) $ and $%
S_{312}^{out}\left( \omega ^{\prime }\right) $ versus the detuning $\Delta $
for (a) $\omega ^{\prime }=0$ and (b) $\omega ^{\prime }=1.0$ by solid,
dashed and dotted line, respectively. The other parameters are the same as
those in Fig. 2.}
\label{fig3}
\end{figure}
In order to study the entanglement properties of output cavity modes, we
need to use quadrature amplitude and phase operators defined by
\begin{eqnarray}
X_{j}^{out} &=&a_{j}^{out}\left( \omega \right) +a_{j}^{\dagger out}\left(
-\omega \right) ,  \nonumber \\
Y_{j}^{out} &=&-i\left[ a_{j}^{out}\left( \omega \right) -a_{j}^{\dagger
out}\left( -\omega \right) \right] ,  \label{10}
\end{eqnarray}
Using Eq. (6) and $X_{j}^{in}=a_{j}^{in}\left( \omega \right)
+a_{j}^{\dagger in}\left( -\omega \right) $, $Y_{j}^{in}=-i\left[
a_{j}^{in}\left( \omega \right) +a_{j}^{\dagger in}\left( -\omega \right)
\right] $, we can obtain the relationships between the input fields and the
output fields as

\smallskip
\begin{eqnarray}
&&
\begin{tabular}{l}
$X_{1}^{out}\left( \omega ^{\prime }\right) =D_{11}X_{1}^{in}\left( \omega
^{\prime }\right) +D_{12}X_{2}^{in}\left( \omega ^{\prime }\right)
+D_{13}X_{3}^{in}\left( \omega ^{\prime }\right) ,$ \\
$Y_{1}^{out}\left( \omega ^{\prime }\right) =D_{11}Y_{1}^{in}\left( \omega
^{\prime }\right) -D_{12}Y_{2}^{in}\left( \omega ^{\prime }\right)
-D_{13}Y_{3}^{in}\left( \omega ^{\prime }\right) ,$%
\end{tabular}
\nonumber \\
&&
\begin{tabular}{l}
$X_{2}^{out}\left( \omega ^{\prime }\right) =D_{21}X_{1}^{in}\left( \omega
^{\prime }\right) +D_{22}X_{2}^{in}\left( \omega ^{\prime }\right)
+D_{23}X_{3}^{in}\left( \omega ^{\prime }\right) ,$ \\
$Y_{2}^{out}\left( \omega ^{\prime }\right) =-D_{21}Y_{1}^{in}\left( \omega
^{\prime }\right) +D_{22}Y_{2}^{in}\left( \omega ^{\prime }\right)
+D_{23}Y_{3}^{in}\left( \omega ^{\prime }\right) ,$%
\end{tabular}
\label{11} \\
&&
\begin{tabular}{l}
$X_{3}^{out}\left( \omega ^{\prime }\right) =D_{31}X_{1}^{in}\left( \omega
^{\prime }\right) +D_{32}X_{2}^{in}\left( \omega ^{\prime }\right)
+D_{33}X_{3}^{in}\left( \omega ^{\prime }\right) ,$ \\
$Y_{3}^{out}\left( \omega ^{\prime }\right) =-D_{31}Y_{1}^{in}\left( \omega
^{\prime }\right) +D_{32}Y_{2}^{in}\left( \omega ^{\prime }\right)
+D_{33}Y_{3}^{in}\left( \omega ^{\prime }\right) ,$%
\end{tabular}
\nonumber
\end{eqnarray}
where we have defined the normalized analyzing frequency $\omega ^{\prime
}=\omega \tau /\kappa $. The explicit expressions for $D_{jk}$ ($j,k$=1-3)
are presented in Appendix C.

The presence of entanglement between the three cavity modes can be
investigated using the sufficient criterion for continuous-variable
tripartite system proposed by van Loock and Furusawa [27]. The sufficient
inseparability criterion for continuous variable tripartite entanglement is
that if any one of the following inequalities is satisfied, genuine
tripartite entanglement is demonstrated. The inequalities are

\begin{eqnarray}
S_{123} &=&V\left[ X_{1}+\left( X_{2}+X_{3}\right) /\sqrt{2}\right] +V\left[
Y_{1}-\left( Y_{2}+Y_{3}\right) /\sqrt{2}\right] <4,  \nonumber \\
S_{231} &=&V\left[ X_{2}+\left( X_{3}+X_{1}\right) /\sqrt{2}\right] +V\left[
Y_{2}-\left( Y_{3}+Y_{1}\right) /\sqrt{2}\right] <4,  \label{12} \\
S_{312} &=&V\left[ X_{3}+\left( X_{1}+X_{2}\right) /\sqrt{2}\right] +V\left[
Y_{3}-\left( Y_{1}+Y_{2}\right) /\sqrt{2}\right] <4,  \nonumber
\end{eqnarray}
where $V\left( A\right) =<A^{2}>-<A>^{2}$. From the above definition, the
correlation spectra of the quadratures of three output cavity fields are
obtained as
\begin{eqnarray}
S_{123}^{out}\left( \omega ^{\prime }\right)  &=&|\sqrt{2}%
D_{11}-D_{21}-D_{31}|^{2}+|\sqrt{2}D_{12}-D_{22}-D_{32}|^{2}+|\sqrt{2}%
D_{13}-D_{23}-D_{33}|^{2},  \nonumber \\
S_{231}^{out}\left( \omega ^{\prime }\right)  &=&\frac{1}{2}\left( |\sqrt{2}%
D_{21}-D_{11}-D_{31}|^{2}+|\sqrt{2}D_{22}-D_{12}-D_{32}|^{2}+|\sqrt{2}%
D_{23}-D_{13}-D_{33}|^{2}\right)   \nonumber \\
&&+\frac{1}{2}\left( |\sqrt{2}D_{21}-D_{11}+D_{31}|^{2}+|\sqrt{2}%
D_{22}-D_{12}+D_{32}|^{2}+|\sqrt{2}D_{23}-D_{13}+D_{33}|^{2}\right) ,
\label{13} \\
S_{312}^{out}\left( \omega ^{\prime }\right)  &=&\frac{1}{2}\left( |\sqrt{2}%
D_{31}-D_{11}-D_{21}|^{2}+|\sqrt{2}D_{32}-D_{12}-D_{22}|^{2}+|\sqrt{2}%
D_{33}-D_{13}-D_{23}|^{2}\right)   \nonumber \\
&&+\frac{1}{2}\left( |\sqrt{2}D_{31}+D_{11}-D_{21}|^{2}+|\sqrt{2}%
D_{32}-D_{12}+D_{22}|^{2}+|\sqrt{2}D_{33}-D_{13}+D_{23}|^{2}\right) .
\nonumber
\end{eqnarray}

The quantum correlations spectra $S_{123}^{out}\left( \omega ^{\prime
}\right) $, $S_{231}^{out}\left( \omega ^{\prime }\right) $ and $%
S_{312}^{out}\left( \omega ^{\prime }\right) $ for three output cavity
fields described in Eq. (13) versus the normalized analyzing frequency $%
\omega ^{\prime }$ are plotted in Fig. $2$ for (a) $\Delta =5$ and (b) $%
\Delta =10$ by solid, dashed and dotted line, respectively. The other
parameters are $\Omega =35$, $g^{2}N=10$, $\gamma =1$ and $\kappa =0.1$. The
satisfaction of one of the three inequalities $S_{123}^{out}\left( \omega
^{\prime }\right) <4$, $S_{231}^{out}\left( \omega ^{\prime }\right) <4$ and
$S_{312}^{out}\left( \omega ^{\prime }\right) <4$ is sufficient to
demonstrate genuine tripartite entanglement. In order to analyze the
entanglement properties of the three cavity modes, we present all three
correlations $S_{ijk}^{out}\left( \omega ^{\prime }\right) $ and find that
the indices of the three cavity modes are crucial. When the cavity modes are
symmetric, the indices of the cavity modes are not important as the three
correlations give the same result. But when the cavity modes are asymmetric,
the indices are crucial in that the three correlations will give different
results. As shown in Fig. 2(a), all three correlations are below 4 in a wide
frequency range thus all three inequalities are satisfied. So the three
output cavity modes are entangled. Among the three correlations, $%
S_{123}^{out}\left( \omega ^{\prime }\right) $ gives the minimum values with
the same parameters. When the inequalities are satisfied, the smaller the
values of correlations are the larger the correlation degree. When we
increase the detuning $\Delta $ to $10$ and keep other parameters unchanged
as shown in Fig. 2(b), correlations $S_{123}^{out}\left( \omega ^{\prime
}\right) $ and $S_{231}^{out}\left( \omega ^{\prime }\right) $ are always
below $4$ in a wide frequency range while the correlation $%
S_{312}^{out}\left( \omega ^{\prime }\right) $ is larger than $4$ in a
frequency zone around the central analyzing frequency $\omega ^{\prime }=0$.
Thus tripartite entanglement is also demonstrated between the three output
cavity modes. Compared with Fig. 2(a), the minimum value of $%
S_{123}^{out}\left( \omega ^{\prime }\right) $ is smaller, which means that
the correlation degree is also increased with the detuning. For both cases,
we also see that the large correlation can be obtained at low analyzing
frequency $\omega ^{\prime }$.

In Fig. $3$, we plot $S_{123}^{out}\left( \omega ^{\prime }\right) $, $%
S_{231}^{out}\left( \omega ^{\prime }\right) $ and $S_{312}^{out}\left(
\omega ^{\prime }\right) $ as a function of detuning $\Delta $ for (a) $%
\omega ^{\prime }=0$ and (b) $\omega ^{\prime }=1.0$ by solid, dashed and
dotted line, respectively. The remain parameters are the same as those in
Fig. 2. We also see that the correlation $S_{123}^{out}\left( \omega
^{\prime }\right) $ gives the minimum values with the same parameters. It is
seen from Fig. 3(a) and 3(b) that, correlations $S_{123}^{out}\left( \omega
^{\prime }\right) $ and $S_{231}^{out}\left( \omega ^{\prime }\right) $
always satisfy the inequalities while $S_{312}^{out}\left( \omega ^{\prime
}\right) $ only satisfy the inequality in a small frequency range. Thus
tripartite entanglement between the three output cavity modes is
demonstrated again. It is worthwhile to point out that when the analyzing
frequency $\omega ^{\prime }=0$, the system reaches its steady state. Thus
at steady state, we can also obtain entangled tripartite light. This is in
contrast with the results in Ref. [20], where the entanglement between the
three cavity modes is time dependent. In that case all the mean photon
numbers are amplified as time increases. Thus, the entanglement for three
cavity modes can not be kept for a long time. And among the three
correlations, $S_{123}^{out}\left( \omega ^{\prime }\right) $ decreases with
the increasing detuning, while correlations $S_{231}^{out}\left( \omega
^{\prime }\right) $ and $S_{312}^{out}\left( \omega ^{\prime }\right) $
first decrease than increase with increasing detuning. So correlation $%
S_{123}^{out}\left( \omega ^{\prime }\right) $ is the best choice when we
investigate the entanglement properties of the three cavity modes. Compared
with Fig. 3(a) and 3(b), we find that the minimal values of correlations in
Fig. 3(a) are smaller than those in Fig. 3(b). This indicates that the
correlation degree is large when the analyzing frequency $\omega ^{\prime }$
is small.

\section{Conclusion}

In conclusion, we have examined the entanglement properties of three cavity
modes interacting with three-level $\Lambda $ atomic system coupled by two
extra classical fields. As the classical fields are stronger than the cavity
fields, we adopt the dressed-atom approach to calculate the equation for the
cavity fields. After tracing out the atomic variables, we obtain the master
equation of the cavity modes and analyze the entanglement properties of the
output fields. The tripartite entanglement of the three output fields is
demonstrated theoretically by a sufficient inseparability criterion and the
entanglement characteristics are presented. This scheme of three-mode
continuous variable entanglement generation using atomic coherence is useful
in quantum information processing.

{\bf Acknowledgments}

This work is supported by the Scientific Research Plan of the Provincial Education Department in Hubei (Grant No. Q20101304) and NSFC under Grant No. 11147153.

{\bf Appendix A}

In this Appendix, we present the atomic decay term in terms of the dressed
atomic states as

\begin{eqnarray}
{\cal L}_{a}\rho &=&\sum_{j,k=0,+,-;j\neq k}\left( {\cal L}_{jk}^{kj}\rho +%
{\cal L}_{ph}^{kj}\rho \right) +\sum_{j,k=+,-;j\neq k}{\cal L}_{in}^{kj}\rho
,  \nonumber \\
{\cal L}_{jk}^{kj}\rho &=&\frac{_{\gamma _{jk}}}{2}\left( 2\sigma
_{p}^{kj}\rho \sigma _{p}^{kj}-\sigma _{p}^{kj}\sigma _{p}^{kj}\rho -\rho
\sigma _{p}^{kj}\sigma _{p}^{kj}\right) ,  \nonumber \\
{\cal L}_{ph}^{kj}\rho &=&\epsilon _{kj}\frac{_{\gamma _{ph}^{kj}}}{4}\left(
2\sigma _{jk}\rho \sigma _{kj}-\sigma _{kj}\sigma _{jk}\rho -\rho \sigma
_{kj}\sigma _{jk}\right) ,  \eqnum{A1} \\
\sigma _{p}^{kj} &=&\sigma _{kk}-\sigma _{jj},  \nonumber \\
{\cal L}_{in}^{kj}\rho &=&\gamma _{c}\left( \sigma _{j0}\rho \sigma
_{k0}+\sigma _{0j}\rho \sigma _{0k}\right) ,  \nonumber
\end{eqnarray}
with $\epsilon _{kj}=1$, for $k,j=0+,0-,+-$, otherwise $\epsilon _{kj}=0$.
The parameters in the above equations are

\begin{eqnarray}
\gamma _{+-} &=&\gamma _{-+}=\frac{\gamma }{4}c^{2}\left( 1+s^{2}\right) ,
\nonumber \\
\gamma _{+0} &=&\gamma _{-0}=\frac{\gamma }{2}s^{2}\left( 1+s^{2}\right) ,
\nonumber \\
\gamma _{0+} &=&\gamma _{0-}=\frac{\gamma }{2}c^{4},\gamma _{c}=\frac{\gamma
}{2}c^{2}s^{2},  \eqnum{A2} \\
\gamma _{ph}^{0+} &=&{\cal \gamma }_{ph}^{0-}=\gamma c^{2}s^{2},\gamma
_{ph}^{+-}=\frac{\gamma }{2}c^{4}.  \nonumber
\end{eqnarray}
{\bf Appendix B}

In this appendix, we present the explicit expressions for the coefficients $%
A_{jk}$ and $B_{jk}$ ($j,k=$1-3) in the equation of motion for the density
operator $\rho _{c\text{ }}$of the cavity modes (Eq. ($4$) ):
\begin{eqnarray}
A_{11} &=&g^{2}N\left( c_{1}e_{1}\rho _{00}^{s}+c_{3}e_{2}\rho
_{++}^{s}\right) \text{, }B_{11}=g^{2}N\left( c_{3}e_{2}\rho
_{00}^{s}+c_{1}e_{1}\rho _{--}^{s}\right) ,  \nonumber \\
A_{22} &=&g^{2}N\left( c_{2}e_{3}\rho _{00}^{s}+c_{3}e_{4}\rho
_{--}^{s}\right) \text{, }B_{22}=g^{2}N\left( c_{2}e_{3}\rho
_{++}^{s}+c_{3}e_{4}\rho _{00}^{s}\right) ,  \nonumber \\
A_{33} &=&g^{2}N\left( c_{1}e_{1}\rho _{00}^{s}+c_{3}e_{2}\rho
_{--}^{s}\right) \text{, }B_{33}=g^{2}N\left( c_{1}e_{1}\tilde{\rho}%
_{++}+c_{3}e_{2}\rho _{00}^{s}\right) ,  \nonumber \\
A_{12} &=&g^{2}N\left( c_{2}e_{2}\rho _{++}^{s}-c_{3}e_{1}\rho
_{00}^{s}\right) \text{, }B_{12}=g^{2}N\left( c_{2}e_{2}\rho
_{00}^{s}-c_{3}e_{1}\rho _{--}^{s}\right) ,  \nonumber \\
A_{13} &=&g^{2}N\left( c_{1}e_{2}\rho _{++}^{s}+c_{3}e_{1}\rho
_{00}^{s}\right) \text{, }B_{13}=g^{2}N\left( c_{1}e_{2}\rho
_{00}^{s}+c_{3}e_{1}\rho _{--}^{s}\right) ,  \eqnum{B1} \\
A_{21} &=&g^{2}N\left( c_{3}e_{3}\rho _{00}^{s}-c_{1}e_{4}\rho
_{--}^{s}\right) \text{, }B_{21}=g^{2}N\left( c_{3}e_{3}\rho
_{++}^{s}-c_{1}e_{42}\rho _{00}^{s}\right) ,  \nonumber \\
A_{23} &=&g^{2}N\left( c_{1}e_{3}\rho _{00}^{s}-c_{3}e_{4}\rho
_{--}^{s}\right) \text{, }B_{23}=g^{2}N\left( c_{1}e_{3}\rho
_{++}^{s}-c_{3}e_{4}\rho _{00}^{s}\right) ,  \nonumber \\
A_{31} &=&g^{2}N\left( c_{3}e_{1}\rho _{00}^{s}+c_{1}e_{2}\rho
_{--}^{s}\right) \text{, }B_{31}=g^{2}N\left( c_{3}e_{1}\rho
_{++}^{s}+c_{1}e_{2}\rho _{00}^{s}\right) ,  \nonumber \\
A_{32} &=&g^{2}N\left( c_{2}e_{1}\rho _{00}^{s}-c_{3}e_{2}\rho
_{--}^{s}\right) \text{, }B_{32}=g^{2}N\left( c_{2}e_{1}\rho
_{++}^{s}-c_{3}e_{2}\rho _{00}^{s}\right) .  \nonumber
\end{eqnarray}
where $e_{1}=\Gamma c_{1}-\gamma _{c}c_{3}$, $e_{2}=\Gamma c_{3}-\gamma
_{c}c_{1}$, $e_{3}=$ $\Gamma c_{2}+\gamma _{c}c_{3}$ and $e_{4}=\Gamma
c_{3}+\gamma _{c}c_{2}$ with $\Gamma =\gamma _{ph}^{0+}+\frac{1}{2}\left(
\gamma _{+-}+\gamma _{+0}+\gamma _{-0}+\gamma _{0+}\right) +\frac{1}{4}%
\left( \gamma _{ph}^{0-}+\gamma _{ph}^{+-}\right) $.

{\bf Appendix C}

In this appendix, we will give the explicit expressions for the coefficients
$D_{jk}$ ($j,k=$1-3) in the relations between the input fields and the
output fields in Eq. (11):
\begin{eqnarray}
D_{11} &=&-1+\chi _{0}\left[ \chi _{22}\chi _{33}-\left(
A_{13}-B_{13}\right) \left( A_{32}-B_{32}\right) \right] ,  \nonumber \\
D_{22} &=&-1+\chi _{0}\left[ \chi _{11}\chi _{33}-\left(
A_{13}-B_{13}\right) \left( A_{31}-B_{31}\right) \right] ,  \nonumber \\
D_{33} &=&-1+\chi _{0}\left[ \chi _{11}\chi _{22}-\left(
A_{12}-B_{12}\right) \left( A_{21}-B_{21}\right) \right] ,  \nonumber \\
D_{12} &=&\chi _{0}\left[ \left( A_{13}-B_{13}\right) \left(
A_{32}-B_{32}\right) -\left( A_{12}-B_{12}\right) \chi _{33}\right] ,
\nonumber \\
D_{13} &=&\chi _{0}\left[ \left( A_{12}-B_{12}\right) \left(
A_{23}-B_{23}\right) -\left( A_{13}-B_{13}\right) \chi _{22}\right] ,
\eqnum{C1} \\
D_{21} &=&\chi _{0}\left[ \left( A_{23}-B_{23}\right) \left(
A_{31}-B_{31}\right) -\left( A_{21}-B_{21}\right) \chi _{33}\right] ,
\nonumber \\
D_{23} &=&\chi _{0}\left[ \left( A_{13}-B_{13}\right) \left(
A_{21}-B_{21}\right) -\left( A_{23}-B_{23}\right) \chi _{11}\right] ,
\nonumber \\
D_{31} &=&\chi _{0}\left[ \left( A_{32}-B_{32}\right) \left(
A_{21}-B_{21}\right) -\left( A_{31}-B_{31}\right) \chi _{22}\right] ,
\nonumber \\
D_{32} &=&\chi _{0}\left[ \left( A_{12}-B_{12}\right) \left(
A_{31}-B_{31}\right) -\left( A_{32}-B_{32}\right) \chi _{11}\right] ,
\nonumber
\end{eqnarray}
where
\begin{eqnarray*}
|D_{0}| &=&\chi _{11}\left[ \chi _{22}\chi _{33}-\left( A_{23}-B_{23}\right)
\left( A_{32}-B_{32}\right) \right] \\
&&+\left( A_{12}-B_{12}\right) \left[ \left( A_{23}-B_{23}\right) \left(
A_{31}-B_{31}\right) -\left( A_{21}-B_{21}\right) \chi _{33}\right] \\
&&+\left( A_{13}-B_{13}\right) \left[ \left( A_{21}-B_{21}\right) \left(
A_{32}-B_{32}\right) -\left( A_{31}-B_{31}\right) \chi _{22}\right] .
\end{eqnarray*}
with the parameters $\chi _{jj}=\kappa \left( \frac{A_{jj}}{\kappa }-\frac{%
B_{jj}}{\kappa }-\frac{1}{2}-i\omega ^{\prime }\right) $ ($j=$1-3) and $\chi
_{0}=-\frac{\kappa }{|D_{0}|}$. And we have used the equation $\kappa
_{1}=\kappa _{2}=\kappa _{3}=\kappa $.

\end{document}